\documentclass[12pt]{article}
\usepackage[utf8]{inputenc}
\usepackage{graphicx}
\usepackage{subfig}
\usepackage{epstopdf}
\usepackage{amssymb,amsmath,epsfig}
\usepackage[left=2cm,right=2cm,top=2cm,bottom=2cm]{geometry}
\usepackage{cite}
\usepackage{amsfonts}
\usepackage{lscape}
\usepackage{accents}
\usepackage{geometry}
\usepackage{latexsym}
\usepackage{afterpage}
\usepackage{amstext}
\usepackage{amscd}
\usepackage{setspace}
\usepackage{url}
\usepackage{anyfontsize}
\usepackage{extsizes}
\usepackage{placeins}

\begin{document}

\title{\bf Development of complexity induced
frameworks for charged cylindrical polytropes}
\author{Shiraz Khan\thanks{f2016265004@umt.edu.pk}, S. A. Mardan\thanks{syedalimardanazmi@yahoo.com}, M. A. Rehman\thanks{aziz.rehman@umt.edu.pk}\\
Department of Mathematics,\\
University of Management and Technology,\\
C-II. Johar Town. Lahore-54590, Pakistan.}
\date{\today}
\maketitle
{
\allowdisplaybreaks
\begin{abstract}
The main theme of this work is the development of complexity induced generalized frameworks for static cylindrical polytropes. We consider two different definitions of generalized polytopes with charged anisotropic inner fluid distribution. A new methodology based on complexity factor for the generation of consistent sets of differential equations will be presented. We conclude our work by carrying out
graphical analysis of developed frameworks.  \\

{\bf Keywords:} Charged anisotropic fluid; Cylindrical symmetry;  Generalized polytropes; Complexity factor.
\end{abstract}

\section{Introduction}
There are many physical parameters which are essential to describe the matter source like anisotropy, dissipation, charge etc. These physical parameters can be characterized as important ingredient in the study of self gravitating sources. Self gravitating body trigger more electromagnetic charge to stable its position in the strong gravitational field.  Bonner \cite{28} showed that in spherical static charged fluid distribution the energy of electric field contributed to the mass of system. He studied a model in which gravitational pull of charged sphere balanced its electrical repulsion. Bondi \cite{29} gave a precise and useful method of examining the contraction in radiating compact star by using the field equations in Minkoweski coordinates. Thorne \cite{50} introduced the idea of C-energy for a system with cylindrical symmetry and discussed many of its properties. In \cite{27}        Bekenstein made a formalism to split the mass of charged black hole into two parts, irreducible part and charged reducible part in spherical symmetry. Mak and Harko \cite{51} found an analytical solution of gravitational field equations for interior of a charged strange quark star.  Takisa and Maharaj \cite{31} used charged spherical anisotropic fluid distribution to find the exact solution of Einstein-Maxwell field equations. They also discussed some graphical plots of different quantities with help of polytropic equation of state (PEoS).  Sharif and Sadiq \cite{34} applied perturbations technique to developed a force distribution function in charged spherical symmetry and found that compact object remained stable in the presence of charge.\\

Polytropes, in the context of general relativity have great importance in study of different changes in characteristics and physical models of a astronomical objects. These characteristics and physical models can easily be described by the PEoS. Polytropes are the solution of Lane-Emden equation (LEe), which is a pair of non linear differential equations so these are always attracted by many researchers in  astrophysics and mathematics. Lane \cite{39} used polytropes to give some basic results associated to the modeling of cosmological structures. Chandrasekhar \cite{1} determined the maximum mass limit for a stable white dwarf star in Newtonian physics by using the idea of polytropes. Tooper \cite{2} analyzed the solution of basic field equations in reference of theory of relativity by making use of PEoS for spherical compressible fluid under gravitational equilibrium. He \cite{3} also found the numerical solutions of hydrostatic equilibrium equation in general relativity using spherical compressible fluid  which was governed by the relation of pressure-energy density. Kaplan and Lupanov \cite{4} studied the relativistic effects in the theory of the structure of polytropic sphere and they  obtained an analytical relation between central  density and spherical mass in weak relativity. For different values of polytropic index, Managhan and Roxburgh \cite{40} investigated the structure of rotating polytropes. For this purpose they used method of approximation for the matching of two solutions at an interface.  Kaufmann \cite{5} used PEoS under static spherical symmetry to obtain a single integro-differential equation and its solution depended on different values of polytropic index $n$. Occhionero \cite{6} evaluated the impact of rotation on the structure of polytropes for $n\geqslant2$,  which were in equilibrium to the second order with suitable parameter. Kovetz \cite{7} reviewed the theory of slowly rotating polytropes of Chandrasekhar \cite{1} and removed some inconsistencies in it.\\
 
 Horedt \cite{8} discussed the instability of weakly distorted polytropic sphere for polytropic index $n>3$. He also observed that slowly rotating cylinder and polytropic rings did not show any instability under the external pressure. Sharma\cite{9} tabulated the values of radius of static polytropic sphere for polytropic index $n=0, 1, 3$ and values of other physical parameters for $n=0,1$ by using the pade $(2,2)$ approximation. Singh and Singh \cite{41} carried out a study of rotationally distorted and tidally polytropes by using the method of \cite{40}. Horedt \cite{11} analyzed the properties like mass acceleration gravitational potential and mean density of $N-$ dimensional radially symmetric polytropes with the help of gamma function. He also discussed the mass$-$radius relation in case of cylindrical and spherical symmetry. Pandey \cite{12} calculated various parameters to study the spherical static structure by using PEoS. For a certain region of sun interior, Hendry \cite{42} developed three polytropic models.   \\
 
  Herrera and Barreto \cite{14} evaluated the relativistic polytropes by using the two different definitions of relativistic polytropes giving same Newtonian limit for self gravitating sphere. They \cite{16} also gave a general structure for the modeling of polytropes in the context of general relativity and derived the LEe. Herrera et al. \cite{17} used the PEoS to analyze the spherically symmetric fluid which is distributed conformally flat and constructed models of highly compact star for anisotropic polytropes in mass density case. Herrera et al. \cite{18} examined the effects of different variations in energy density and pressure anisotropy under cracking technique for spherical non static compact objects satisfying the PEoS.\\
  
  The generalized polytropic equation of state (GPEoS) is used to discuss the generalized polytrops (GPs) for the study of gravitating objects. This equation consists of two parts: $(i)$ polytropic part $P_{r}= K\mu^{\gamma}_{o}=K\mu_{o}^{1+\frac{1}{n}}$ and $(ii)$ linear part  $P_{r}=\alpha_{1}\mu_{o}$. Combination of these two parts define the GPEoS \cite{19} as 
 \begin{equation}\label{1}
 P_{r}= K\mu^{\gamma}_{o}+\alpha_{1}\mu_{o}=K\mu_{o}^{1+\frac{1}{n}}+\alpha_{1}\mu_{o} ,
 \end{equation}
 where $P_{r}$, $K$, $ \gamma$, $\alpha_{1}$ and $n$ are called principal stress, polytropic constant , polytropic exponent, linear coefficient constant and polytropic index respectively. Change of   $\mu_{o}$ by $\mu$ gives 

  \begin{equation}\label{2}
   P_{r}= K\mu^{1+\frac{1}{n}}+\alpha_{1}\mu .
 \end{equation}\\
Azam et al. \cite{19, 35} carried out spherical and cylindrical symmetric fluid distribution to study the charged polytropes with relativistic GPEoS. Mardan et al.\cite{43,44} used spherical symmetric GPs to investigate some gravitating objects. They found exact solutions of field equations by taking different values of polytropic index $n$ and analyzed some mathematical models which were found physically viable and well behaved. Mardan et al. \cite{45,46} established new classes of polytropic models using spherical symmetry and analyzed the mass and radius of different astronomical objects.\\

Herrera \cite{21} used orthogonal splitting of  Riemann tensor into structure scalars for a self gravitating system to present a new concept of complexity factor (CF). Abbas and Nazar \cite{47} carried out this concept of vanishing CF for self gravitating object in the context of modify gravity. Sharif and Butt \cite{36,48} implemented the CF on cylindrical system and discussed the electromagnetic effect on this system. Khan et al. \cite{32,49} used the idea of CF to study the GPs and charged GPs for spherical self gravitating fluid.  They \cite{52} also studied the same idea for static cylindrical GPs with CF.\\
 
The layout of this paper will be follow. Section $\textbf{2}$ will contain the detail of basic field equations using cylindrical static symmetry and  Tolman-Opphenheimer-Volkoff (TOV) equation under electromagnetic effect. In section $\textbf{3}$, we will use the the Weyl tensor to discuss the mass function for self gravitating object. In section $\textbf{4}$ a discussion will be carried out about charged GPs for two cases $(i)$ mass density and $(ii)$ energy density. This section also contains physical conditions about these two cases. Section $\textbf{5}$  will be devoted for the study of CF which is defined through structure scalars obtained by orthogonal splitting of Riemann tensor. We will give a graphical solution of cylindrical charged GPs with CF in section $\textbf{6}$. In last section $\textbf{7}$ we will conclude our work. 
\section{Basic equations}
 Let us consider a static cylindrical symmetric line element, as
\begin{equation}\label{3}
ds^2=-A^{2}dt^{2}+B^{2}dr^{2}+C^{2} d\theta^{2}+a^{2} C^{2}d z^{2},
\end{equation}
where $A=A(r)$, $B=B(r)$, $C=C(r)$, a is an arbitrary constant and coordinates are: $x^{0}=t$, $x^{1}=r$, $x^{2}=\theta$, $x^{3}=z$. The electromagnetic tensor is defined by
\begin{equation*}
\Gamma^{i}_{j}=\frac{1}{4\pi}(-F^{i}_{\mu}F^{\mu}_{j}+\frac{1}{4}F^{\mu\nu}F_{\mu\nu}g^{i}_{j}),
\end{equation*}
where $F_{ij}=\varphi_{j,i}-\varphi_{i,j}$ is called the Maxwell tensor and $\varphi_{j}$ denotes four potential given by $\varphi_{i}=\varphi\delta^{0}_{i}$. The Maxwell equations are 
\begin{equation*}
F^{ij}_{;j}=\phi_{o}J^{i}, \quad F_{[ij;k]}=0,
\end{equation*}
where $\phi_{o}$ and $J^{i}=\sigma u^{i}$ represents the magnetic permeability and the four current respectively, where $\sigma$ is the charge density. The Maxwell equation for metric $(3)$, is
\begin{equation*}
\varphi^{'}\Big(\frac{2}{r}-\frac{A^{'}}{A}-\frac{B^{'}}{B}\Big)+\varphi^{''}=4\pi \sigma AB^{2},
\end{equation*}
it implies that
\begin{equation*}
r \varphi^{'}=A B q(r),
\end{equation*}
where $q(r)=4\pi\int_{0}^{r}\sigma B \overline{r} d\overline{r}$ denote the total charge within the cylinder .\\
 The stress-energy tensor is defined as
\begin{equation}\label{4}
T^{\lambda}_{\gamma}= \Delta^{\lambda}_{\gamma}+Ph^{\lambda}_{\gamma}+ \mu u^{\lambda}u_{\gamma}+\Gamma^{\lambda}_{\gamma},
\end{equation}
with
\begin{eqnarray}
&&\Delta^{\lambda}_{\gamma}=\frac{\Delta}{3}(h^{\lambda}_{\gamma}+3 s^{\lambda}s_{\gamma}); \; 3 P=P_{r}+2 P_{\perp}\notag\\
&& h^{\lambda}_{\gamma}=\delta^{\lambda}_{\gamma}-u^{\lambda}u_{\gamma};\quad \Delta=-(P_{\perp}-P_{r}).
\end{eqnarray}
here $ P_{\perp}$ is an other principal stress, four vector and four velocity respectively are defined as
 $s_{\mu}=(0,B^{-1},0,0)$ and $u_{\mu}=(A^{-1},0,0,0)$, 
with properties:
\begin{center}
$ s^{\mu}u_{\mu}=0,\quad s^{\mu}s_{\mu}=1,\quad u^{\mu}u_{\mu}=-1$.
\end{center}
Basic equations are
\begin{equation}\label{5}
8\pi\mu=\frac{2C^{**}}{B^{2}C}-\frac{2B^{*}C^{*}}{B^{3}C}+\frac{C^{*2}}{B^{2}C^{2}}-\frac{q^{2}}{r^{4}},
\end{equation}
\begin{equation}\label{6}
-8\pi P_{r}=\frac{C^{*2}}{B^{2}C^{2}}+\frac{2 A^{*} C^{*}}{A B^{2}C}+\frac{q^{2}}{r^{4}},
\end{equation}
\begin{equation}\label{7}
-8\pi P_{\perp}=\frac{A^{*}C^{*}}{AB^{2}C}+\frac{C^{**}}{B^{2}C}-\frac{{B^{*}C^{*}}}{B^{3}C}+\frac{A^{**}}{A B^{2}}-\frac{A^{*} B^{*}}{A B^{3}}-\frac{q^{2}}{r^{4}},
\end{equation}
where $`*'$ indicates the derivative w. r. t. $`r'$. The corresponding exterior geometry is considered as \cite{37}.
\begin{equation}\label{8}
ds^{2}=\frac{2M}{R}d\nu^{2}-2dRd\nu+R^{2}(d\theta^{2}+a^{2})dz^{2},
\end{equation}
where $M$ represents the total mass in the exterior. On the hyper surface $\Sigma$, the necessary and sufficient conditions for smooth matching of two metrics $(\ref{3})$ and $(\ref{8})$ are given in \cite{37} 
 as $C(r)=r$ $\Rightarrow$ $C^{*}(r)=1$ and $C^{**}(r)=0$, then Eqs. $(\ref{5}-\ref{7})$ become,
\begin{equation}\label{9}
\mu=\frac{1}{8\pi}[-\frac{2B^{*}}{B^{3}r}+\frac{1}{B^{2}r^{2}}-\frac{q^{2}}{r^{4}}],
\end{equation}
\begin{equation}\label{10}
P_{r}=-\frac{1}{8\pi}[\frac{1}{B^{2}r^{2}}+\frac{2 A^{*}}{A B^{2}r}+\frac{q^{2}}{r^{4}}],
\end{equation}
\begin{equation}\label{11}
P_{\perp}=-\frac{1}{8\pi}\big[\frac{A^{**}}{A B^{2}}-\frac{A^{*} B^{*}}{A B^{3}}+\frac{A^{*}}{AB^{2}r}-\frac{{B^{*}}}{B^{3}r}-\frac{q^{2}}{r^{4}}\big].
\end{equation}
Solving Eqs. $(\ref{9}-\ref{11})$ simultaneously we obtain generalized TOV equation as
\begin{equation}\label{12}
P^{*}_{r}=-\frac{A^{*}}{A}(\mu+P_{r})+\frac{2(P_{\bot}-P_{r})}{r}+\frac{qq^{*}}{4\pi r^{4}}.
\end{equation}
 \section{Mass function and Weyl tensor}
Cylindrical energy (Or C-energy) \cite{50} is to be considered as a very useful tool to analyze the finite and infinite cylindrical system. It is defined in the form of mass function as
\begin{equation}\label{13}
\textbf{E}=\frac{1}{8}(1-\frac{1}{l}\nabla_{\rho}r\nabla^{\rho}r)+\frac{q^{2}}{2C},
\end{equation}
it yields
\begin{equation}\label{14}
m(r)\cong \textbf{E} =l\textbf{E}=-\frac{ra}{2B^2}(1-\frac{B^2}{4})+\frac{q^{2}}{2C},
\end{equation}
using Eqs. $(\ref{9})$ and $(\ref{14})$
\begin{equation}\label{15}
m(r)=\frac{ar}{8}-4\pi\int^{r}_{0}\overline{r}^{2}\mu d\overline{r}+\frac{q^{2}}{2r}-\frac{a}{2}\int^{r}_{0}\frac{q^{2}}{\overline{r}^{2}}d\overline{r}.
\end{equation}
From Eqs. $(\ref{9}-\ref{11})$ and $(\ref{14})$, we have
\begin{eqnarray}\label{16}\notag
&&m=\frac{ra}{8}+8\pi (\mu-P_{r}+P_{\bot})-\frac{1}{B^{2}}(\frac{ra}{2}+\frac{1}{r_{2}})\notag\\
&& \;\;\quad+\frac{1}{AB^{2}}[A^{* *}
-\frac{A^{*}B^{*}}{B}-\frac{A^{*}}{r}\frac{A}{r^{2}}+\frac{A B^{*}}{r B}+\frac{q^{2}}{2 r}+\frac{3q^{2}}{r^{4}}].
\end{eqnarray}
Now with the help of Weyl tensor, we can simplify above expression.
For cylindrical symmetric fluid distribution Wely tensor has  electric and magnetic components. For the purpose  of simplification, we take electric component as
\begin{equation} \label{17}
E_{\alpha\beta}=C_{\alpha\gamma\beta\delta}u^{\gamma}u^{\delta},
\end{equation}
where
\begin{equation}\label{18}
C_{\mu\nu\kappa\lambda}=(g_{\mu\nu\alpha\beta}g_{\kappa\lambda\gamma\delta}-\eta_{\mu\nu\alpha\beta}\eta_{\kappa\lambda\gamma\delta})u^{\alpha}u^{\gamma}E^{\beta\delta},
\end{equation}
with  \qquad\qquad\quad $g_{\mu\nu\alpha\beta}=g_{\mu\alpha}g_{\nu\alpha}$ and $\eta_{\mu\nu\alpha\beta}$. Note that 
\begin{equation}\label{19}
 E_{\alpha\beta}=E(s_{\alpha}s_{\beta}+\frac{1}{3}h_{\alpha\beta}),
 \end{equation}
 and
 \begin{equation}\label{20}
 E=A^{**}-\frac{A^{*}B^{*}}{B}-\frac{A^{*}}{r}-\frac{A}{r^{2}}+\frac{A B^{*}}{r B},
 \end{equation}
  satisfying the following properties:
  \begin{equation}\label{21}
   E_{11}=\frac{1}{3A}E,\; E_{22}=-\frac{r^{2}}{6AB^{2}}E,\; E_{33}=\frac{a^{2}{r^{2}}}{6AB^{2}}E.
\end{equation}
From Eqs. $(\ref{16})$ and $(\ref{20})$ we have
\begin{equation}\label{22}
m=\frac{ra}{8}+8\pi (\mu-P_{r}+P_{\bot})-\frac{1}{B^{2}}(\frac{ra}{2}+\frac{1}{r^{2}})+\frac{E}{AB^{2}}+\frac{q^{2}}{2 r}+\frac{3q^{2}}{r^{4}}.
\end{equation} 
Using Eqs.  $(\ref{15})$ in $(\ref{22})$ 
\begin{eqnarray}\label{23}\notag
&&E=\frac{4}{3}\pi a A B^{2}\int^{r}_{0}\overline{r}^{3}\mu^{*} d \overline{r}-\frac{4}{3}\pi a A B^{2}r^{3}\mu+A(\frac{ra}{2}+\frac{1}{r^{2}})\notag\\
&&\quad-8\pi A B^{2}\mu +8 \pi A B^{2}(P_{r}-P_{\bot})-3 A B^{2}\frac{q^{2}}{r^{4}}-AB^{2}\frac{a}{2}\int_{0}^{r}\frac{q^{2}}{\overline{r}^{2}}d\overline{r}.
\end{eqnarray}
Then Eq. $(\ref{22})$ will be 
\begin{equation}\label{24}
m=\frac{ra}{8}+ \frac{4}{3}\pi a\int^{r}_{0}\overline{r}^{3}\mu^{*} d \overline{r}-\frac{4}{3}\pi a r^{3}\mu+\frac{q^{2}}{2 r}-\frac{a}{2}\int_{0}^{r}\frac{q^{2}}{\overline{r}^{2}}d\overline{r}.
\end{equation}
 Using Eq. $(\ref{14})$ in $(\ref{10})$ 
\begin{equation}\label{25}
\frac{A^{*}}{A}=-\frac{2a(8\pi r^{3}P_{r}-\frac{q^{2}}{r})}{r^{2}a-8rm+4q^{2}}-\frac{1}{2r}.
\end{equation}
Put Eq. $(\ref{25})$ in $(\ref{10})$, the TOV equation becomes
\begin{equation}\label{26}
P^{*}_{r}=\Big(\frac{2a(8\pi r^{3}P_{r}-\frac{q^{2}}{r})}{r^{2}a-8rm+4q^{2}}+\frac{1}{2r}\Big)(\mu+P_{r})+\frac{2(P_{\bot}-P_{r})}{r}.
\end{equation}
   \section{Charged generalized cylindrical polytropes }
Since GPEoS are widely used $[27-33,37,38,39]$ to discuss the different characteristics of inner structure of self gravitating objects. So we use it for cylindrical static symmetry anisotropic fluid in the presence of charge as  
 \subsection{Case 1}
  
 \begin{equation}\label{39}
   P_{r}=K\mu_{o}^{1+\frac{1}{n}}+\alpha_{1}\mu_{o},
 \end{equation}
energy density $\mu$ connected with the mass density $\mu_{o}$ \cite{17} as
     \begin{equation}\label{40}
     \mu=\mu_{o}+nP_{r}.
     \end{equation}
 following assumptions are to be considered
     \begin{equation}\label{41}
     r=\frac{\xi}{N}, \quad \alpha=\frac{P_{rc}}{\mu_{c}} ,\quad N^{2}=\frac{4\pi\mu_{c}}{(1+n)\alpha}.
     \end{equation}
     \begin{equation}\label{42}
     \psi_{o}=\frac{\mu_{o}}{\mu_{oc}},\quad v(\xi)=\frac{m(r)N^{3}}{4\pi\mu_{c}}.
     \end{equation}
     Then TOV Eq. $(\ref{13})$ becomes
     
\begin{eqnarray}\label{43}\notag
&&\frac{-1}{\alpha ^2 N^4}\big[4 a P_{rc} {\psi_{ o}}^n ((n+1) {\psi_{ o}}(\alpha -{\alpha_{ 1}}+\alpha  {\alpha_{ 1}} n)-(\alpha  n-1)\notag\\&& ({\alpha_{ 1}}+{\alpha_{ 1}} n+1)) (4 \pi  \xi ^4 {P_{rc}} {\psi_{ o}}^n ({\alpha_{ 1}}+{\psi_{ o}}(\alpha -{\alpha_{ 1}}+\alpha  {\alpha_{ 1}} n)\notag\\
&&+\alpha  {\alpha_{ 1}} (-n))-\alpha  N^4 q^2)\big]+\frac{(\alpha  N^2 \xi -32 \pi  {P_{rc}} v)\xi}{\alpha  N^4}
+4 q^2(-\frac{N^3 q q'}{2 \pi  \xi ^3}\notag\\
&&-4 \Delta+\alpha^{-1} \big[2 \xi  {P_{rc}} {\psi_{ o}}^{n-1} {\psi_{ o}}' ((n+1) {\psi_{ o}}(\alpha -{\alpha_{ 1}}+\alpha  {\alpha_{ 1}} n)\notag\\
&&+{\alpha_{ 1}} n (1-\alpha  n))\big].
\end{eqnarray}
 where prime indicates the derivative w. r. t. $\xi$. Using Eqs. $(\ref{41}, \ref{42})$ in (\ref{15}) we have
     \begin{eqnarray}\label{44}
     &&\dfrac{dv}{d\xi}=\frac{a}{8(n+1)\alpha}-a\xi^{2}\psi_{o}^{n}\Big[(1-n\alpha)(1+n\alpha_{1})+n(\alpha-\alpha_{1}+n\alpha\alpha_{1})\Big]\notag\\
     &&\qquad+\frac{N}{\alpha(n+1)}\big(\frac{q q'}{\xi}-\frac{N(1+a)q^{2}}{2\xi^{2}}\big).     
 \end{eqnarray}
     At boundary  surface $\xi=\xi_{n}$ such that $\psi_{o}(\xi_{n})=0$ and we have boundary conditions as
     \begin{equation}\label{45}
 \psi_{o}(\xi=0)=1 \quad  and \quad v(\xi=0)=0.
      \end{equation}
      Eqs. $(\ref{43},\; \ref{44})$ constitute the LEe for this case
\begin{eqnarray}\label{46} \notag
&&\frac{1}{2 \pi  \alpha ^2 N^4 \xi ^4}\big[-8 \pi  a \xi ^4 {P_{rc}} {\psi_{o}}^n ((n+1) {\psi_{o}} \beta+(1-\alpha  n) ({\alpha_{1}}+{\alpha_{1}} n+1))\notag\\
&& (-2 \alpha  N^4 q q'-4 \pi  {\alpha_{1}} n \xi ^4 {P_{rc}} (\alpha  n-1) {\psi_{o}}^{n-1} {\psi_{o}}'+4 \pi  \xi ^3 {P_{rc}} {\psi_{o}}^n ((n+1) \xi  \beta {\psi_{o}}'\notag\\
&&-4 {\alpha_{1}} (\alpha  n-1))+16 \pi  \xi ^3 {P_{rc}} \beta {\psi_{o}}^{n+1})+8 \pi  a n \xi ^4 {P_{rc}} {\psi_{o}}^{n-1} {\psi_{o}}' ((n+1) {\psi_{o}} \beta\notag\\
&&+(1-\alpha  n) ({\alpha_{1}}+{\alpha_{1}} n+1)) (\alpha  N^4 q^2-4 \pi  \xi ^4 {P_{rc}} {\psi_{o}}^n ({\alpha_{1}}+{\psi_{o}} \beta-\alpha  {\alpha_{1}} n))\notag\\
&&+8 \pi  a (n+1) \xi ^4 {P_{rc}} \beta {\psi_{o}}^n {\psi_{o}}' (\alpha  N^4 q^2-4 \pi  \xi ^4 {P_{rc}} {\psi_{o}}^n ({\alpha_{1}}+{\psi_{o}} \beta-\alpha  {\alpha_{1}} n))\notag\\
&&-\xi  (2 \alpha  N^2 (4 N^2 q q'+\xi )-32 \pi  {P_{rc}} (\xi  v'+v)) (8 \pi  \alpha  \xi ^3 \Delta +\alpha  N^3 q q'\notag\\
&&-4 \pi  \xi ^4 {P_{rc}} {\psi_{o}}^{n-1} {\psi_{o}}' ((n+1) {\psi_{o}} \beta+{\alpha_{1}} n (1-\alpha  n)))+(4 \alpha  N^4 q^2+\xi  (\alpha  N^2 \xi\notag\\
&& -32 \pi  {P_{rc}} v)) (-\alpha  \xi  (N^3 q'^2+8 \pi  \xi ^3 \Delta ')+\alpha  N^3 q (3 q'-\xi  q'')+4 \pi  \xi ^4 {P_{rc}} {\psi_{o}}^{n-2}\notag\\
&& ({\alpha_{1}} (1-n) n \xi  (\alpha  n-1) {\psi_{o}}'^2+(n+1) {\psi_{o}}^2 \beta (\xi  {\psi_{o}}''+{\psi_{o}}')+n {\psi_{o}} ({\alpha_{1}} \xi \notag\\
&& (1-\alpha  n) {\psi_{o}}''+{\psi_{o}}' ({\alpha_{1}}+(n+1) \xi  \beta {\psi_{o}}'-\alpha  {\alpha_{1}} n))))\big],
\end{eqnarray}
where $\beta=(\alpha -{\alpha_{ 1}}+\alpha  {\alpha_{ 1}} n)$.  

\subsection{ Case 2}

    Now we consider \cite{19} 
     \begin{equation}\label{47}
        P_{r}=K\mu^{1+\frac{1}{n}}+\alpha_{1}\mu,
       \end{equation}
       in this case energy density $\mu$ and mass density $\mu_{o}$ are expressed as \cite{15}
       \begin{equation}\label{48}
        \mu (K\mu_{o}^{1/n}-1)^{n}=(-1)^{n}\mu_{o}.
        \end{equation}
By taking $\psi^{n}=\frac{\mu}{\mu}_{o}$ TOV equation becomes
\begin{eqnarray}\label{49} \notag
&&(\frac{\xi  (\alpha  N^2 \xi -32 \pi  {P_{rc}} v)}{\alpha  N^4}+4 q^2) (-\frac{N^3 q q'}{2 \pi  \xi ^3}-4 \Delta +\frac{1}{\alpha}\big[2 \xi  {P_{rc}} \psi ^{n-1} \psi'\notag\\  
&&((n+1) (\alpha -{\alpha_{1}}) \psi +{\alpha_{1}} n)\big]{\alpha })-\frac{1}{\alpha ^2 N^4}\big[4 a {P_{rc}} \psi ^n ((\alpha -{\alpha_{1}}) \psi\notag\\
&& +{\alpha_{1}}+1) (4 \pi  \xi ^4 {P_{rc}} \psi ^n ((\alpha -{\alpha_{1}}) \psi +{\alpha_{1}})-\alpha  N^4 q^2)\big]=0.
\end{eqnarray} 
and from Eq. $(\ref{15})$ we have
\begin{equation}\label{50}
\dfrac{dv}{d\xi}=\frac{a}{8(n+1)\alpha}-a\xi^{2}\psi^{n}\frac{N}{\alpha(n+1)}\big[(1+a)\frac{q^{2}}{2\xi^{2}}-\frac{qq'}{\xi}\big].
\end{equation}
 Eqs. $(\ref{49}, \ref{50})$ together give the generalized LEe
\begin{eqnarray}\label{51} \notag
&&\frac{-1}{2 \pi  \alpha ^2 N^4 \xi ^4}\Big[16 \pi  a \xi ^4 {P_{rc}} \psi ^{n-1} (({\alpha_{1}}-\alpha ) \psi -{\alpha_{1}}-1) (\alpha  N^4 \psi  q q'\notag\\
&&-2 \pi  \xi ^3 {P_{rc}} \psi ^n (4 \psi  ((\alpha -{\alpha_{1}}) \psi +{\alpha_{1}})+\xi  \psi ' ((n+1) (\alpha -{\alpha_{1}}) \psi\notag\\
&& +{\alpha_{1}} n)))-8 \pi  a n \xi ^4 {P_{rc}} \psi ^{n-1} \psi ' ((\alpha -{\alpha_{1}}) \psi +{\alpha_{1}}+1) (\alpha  N^4 q^2\notag\\
&&-4 \pi  \xi ^4 {P_{rc}} \psi ^n ((\alpha -{\alpha_{1}}) \psi +{\alpha_{1}}))-8 \pi  a \xi ^4 {P_{rc}} (\alpha -{\alpha_{1}}) \psi ^n \psi '\notag\\
&& (\alpha  N^4 q^2-4 \pi  \xi ^4 {P_{rc}} \psi ^n ((\alpha -{\alpha_{1}}) \psi +{\alpha_{1}}))+\xi  (2 \alpha  N^2 (4 N^2 q q'\notag\\
&&+\xi )-32 \pi  {P_{rc}} (\xi  v'+v)) (8 \pi  \alpha  \xi ^3 \Delta +\alpha  N^3 q q'-4 \pi  \xi ^4 {P_{rc}} \psi ^{n-1} \psi '\notag\\
&& ((n+1) (\alpha -{\alpha_{1}}) \psi +{\alpha_{1}} n))+\frac{1}{\psi ^2}\big[(4 \alpha  N^4 q^2+\xi  (\alpha  N^2 \xi -32 \pi  {P_{rc}} v))\notag\\
&& (\alpha  \xi  \psi ^2 (N^3 q'^2+8 \pi  \xi ^3 \Delta' )+\alpha  N^3 \psi ^2 q (\xi  q''-3 q')-4 \pi  \xi ^4 {P_{rc}} \psi ^n ((n+1)\notag\\
&& (\alpha -{\alpha_{1}}) \psi ^2 (\xi  \psi ''+\psi ')+n \psi  ({\alpha_{1}} \xi  \psi ''+\psi ' ({\alpha_{1}}+(n+1) \xi  (\alpha -{\alpha_{1}}) \psi '))\notag\\
&&+{\alpha_{1}} (n-1) n \xi  \psi '^2))\big]\Big]=0.
\end{eqnarray}

Both cases have to  satisfy the physical conditions 
\begin{equation}\label{52}
\mu+\frac{q^2}{8\pi r^4}>0,\quad \frac{P_{r}}{\mu}\leq1+\frac{q^2}{4\pi r^4}, \quad \frac{P_{\perp}}{\mu}\leq1.
\end{equation}
 Conditions $(\ref{52})$ takes the form for case \textbf{1} as 
 \begin{eqnarray}\label{53} \notag
 && (1-n\alpha)+(1+n\alpha_{1})+n(\alpha-\alpha_{1}+n\alpha\alpha_{1})\psi_{o}+\frac{2\pi q^{2}\mu_{c}}{(1+n)^{2}\alpha^{2}\psi_{o}^{n}\xi^{4}}>0,\notag\\
 &&\frac{\alpha_{1}}{\psi_{o}}+(\alpha_{1}+\alpha\psi_{o})\leq [(1-n\alpha)+(1+n\alpha_{1})+(\alpha n \alpha_{1}-\alpha_{1}+\alpha)n\psi_{o}]\notag\\
&&[1+\frac{4\pi q^{2}\mu_{o}^{2}}{(1+n)^{2}\alpha^{2}\xi^{4}}],\;\frac{v}{\xi^{3}\psi_{o}^{n}}-\frac{a}{8\alpha(n+1)\xi^{2}\psi_{o}^{n}}+a\alpha \psi_{o}+a\alpha_{1}(1-n\alpha)\notag\\
&&(1-\psi_{o})+\frac{(3a-1)\pi q^{2}\mu_{c}}{(1+n)^{2}\alpha^{2}\xi^{4}\psi_{o}^{n}}\leq(a-1)[(1-n\alpha)+(1+n\alpha_{1})+(\alpha n \alpha_{1}\notag\\
&&-\alpha_{1}+\alpha)n\psi_{o}]
.
 \end{eqnarray}

 and for Case \textbf{2} these physical conditions $(\ref{52})$ becomes as
  \begin{eqnarray}\label{54}\notag
 &&\psi(1+n)^{2}\alpha^{2}\xi^{2}+2\pi q^{2}\mu_{c}>0,\; \alpha_{1}(1-\psi)+\alpha\psi\leq1+\frac{4\pi q^{2}\mu_{c}}{(1+n)^{2}\alpha^{2}\xi^{4}\psi^{n}},\;\notag\\
 &&\frac{v}{\xi^{3}\psi^{n}}-\frac{a}{8\alpha(n+1)\xi^{2}\psi^{n}}+a\alpha\psi-a\alpha_{1}(1-n\alpha)(1-\psi)\notag\\
 &&\leq \frac{2(1-3 a)\pi q^{2}\mu_{c}}{(1+n)^{2}\alpha^{2}\xi^{4}\psi^{n}}. 
 \end{eqnarray}

\section{Complexity factor}
CF is defined \cite{21} through structure scalars, which are obtained from orthogonal splitting of curvature tensor \cite{24}. This
splitting of curvature tensor give the following tensors \cite{26,33}.
 \begin{equation}\label{27}
  Y_{\alpha\beta}=R_{\alpha\gamma\beta\delta}u^{\gamma}u^{\delta},
  \end{equation}
 \begin{equation}\label{28}
 X_{\alpha\beta}=^{\bullet}R^{\bullet}_{\alpha\gamma\beta\delta}u^{\gamma}u^{\delta}=\frac{1}{2}\eta_{\alpha\gamma}^{\epsilon\mu}R^{\bullet}_{\epsilon\mu\beta\delta}u^{\gamma}u^{\delta},
 \end{equation}
 where $\bullet$ denote the dual tensor i.e. $R^{\bullet}_{\alpha\beta\gamma\delta}=\frac{1}{2}\eta_{\epsilon\mu\gamma\delta}\; R_{\alpha\beta}^{\epsilon\mu}$. While the trace free parts $(Y_{TF},X_{TF})$ and trace part $(Y_{T},X_{T})$ of these tensors are related as \cite{26,33}. and so the tensors in Eqs. $(\ref{27})$ and $(\ref{28})$ can be represented as 
  \begin{equation}\label{29}
   Y_{\alpha\beta}=Y_{TF}(\frac{1}{3}h_{\alpha\beta}+s_{\alpha}s_{\beta})+\frac{1}{3}Y_{T}h_{\alpha\beta},
   \end{equation}
   
   \begin{equation}\label{30}
   X_{\alpha\beta}=X_{TF}(\frac{1}{3}h_{\alpha\beta}+s_{\alpha}s_{\beta})+\frac{1}{3}X_{T}h_{\alpha\beta}.
   \end{equation}
 
  After using the field equations, we have 
   \begin{equation}\label{31}
   X_{T}=8\pi\mu+\frac{q^{2}}{r^{4}},
   \end{equation}
   \begin{equation}\label{32}
   X_{TF}=4\pi\Delta_{\alpha\beta} -E+\frac{q^{2}}{r^{4}}.
   \end{equation}
   Using Eq. $(\ref{23})$ in $(\ref{32})$
   \begin{eqnarray}\notag\label{33}
  && X_{TF}=4\pi\Delta(1-2AB^{2})-\frac{4}{3}\pi a A B^{2}\int^{r}_{0}\overline{r}^{3}\mu^{*} d \overline{r}\notag\\
  &&\qquad+\frac{4}{3}\pi a A B^{2}r^{3}\mu-A(\frac{ra}{2}+\frac{1}{r^{2}})+8\pi A B^{2}\mu\notag\\
  &&\qquad+(1+3 A B^{2})\frac{q^{2}}{r^{4}}+AB^{2}\frac{a}{2}\int_{0}^{r}\frac{q^{2}}{\overline{r}^{2}}d\overline{r},
   \end{eqnarray}
   \begin{equation}\label{34}
   Y_{T}=4\pi(\mu+3P_{r}-2\Delta)+\frac{q^{2}}{r^{4}},
    \end{equation}
   \begin{equation}\label{35}
   Y_{TF}=4\pi\Delta+E+\frac{q^{2}}{r^{4}},
   \end{equation}
   or using Eq. $(\ref{23})$ in $(\ref{35})$
   \begin{eqnarray}\notag\label{36}
   &&Y_{TF}=4\pi\Delta(1+2AB^{2})+\frac{4}{3}\pi a A B^{2}\int^{r}_{0}\overline{r}^{3}\mu^{*} d \overline{r}\notag\\
  &&\qquad-\frac{4}{3}\pi a A B^{2}r^{3}\mu+A(\frac{ra}{2}+\frac{1}{r^{2}})-8\pi A B^{2}\mu\notag\\
  &&\qquad+(1-3 A B^{2})\frac{q^{2}}{r^{4}}-AB^{2}\frac{a}{2}\int_{0}^{r}\frac{q^{2}}{\overline{r}^{2}}d\overline{r}.
   \end{eqnarray}
   From Eqs. $(\ref{33})$ and $(\ref{36})$ 
   \begin{equation}\label{37}
   8\pi\Delta+\frac{2q}{r^{4}}=X_{TF}+Y_{TF}.
   \end{equation}
   
In order to discuss stellar structure the basic field Eqs. ($\ref{9}$\;-\;$\ref{11}$) form a system of three ordinary differential equations (DEs) for static cylindrical symmetry in five variables $(A, B,\mu,P_{\perp},P_{r})$. So we implement condition $Y_{TF}=0$ which gives the vanishing CF Eq. $(\ref{36})$, as \cite{36}
   \begin{eqnarray}\label{38}
  && \Delta=\frac{-1}{4\pi(1+2AB^{2})}\Big[\frac{4}{3}\pi a A B^{2}\int^{r}_{0}\overline{r}^{3}\mu^{*} d \overline{r}
  -\frac{4}{3}\pi a A B^{2}r^{3}\mu\notag\\
  &&\qquad\qquad\qquad\qquad\quad+A(\frac{ra}{2}+\frac{1}{r^{2}})-8\pi A B^{2}\mu\notag\\
  &&\qquad\qquad\qquad\qquad\quad+(1-3 A B^{2})\frac{q^{2}}{r^{4}}-AB^{2}\frac{a}{2}\int_{0}^{r}\frac{q^{2}}{\overline{r}^{2}}d\overline{r}\Big].
   \end{eqnarray}
  
\section{ Vanishing complexity factor with charged generalized cylindrical polytropes} 
As we already have discussed CF as a single scalar and now it is merged with cylindrical GPEoS so that we are able to develop a consistent system of DEs for both cases.
\subsection{Case No.1}
In this case Eqs. $(\ref{41}, \ref{42})$ are used  with $Y_{TF}=0$, as  
\begin{eqnarray}\label{55} \notag
&&4 \pi  (\Delta'+2 B^2 (\Delta A'+A \Delta')+4 A B \Delta B')-\frac{1}{24 N^3 \xi ^5}\Big[24 N^2 q^2 (2 A B\notag\\
&& (B (\frac{\pi  (a+1) \xi ^3 P_{rc}}{\alpha ^2 (n+1)}+6 N^5)-3 N^5 \xi  B')+N^5 (4-3 \xi  B^2 A'))\notag\\
&&+\frac{1}{\alpha ^2}\big[\xi ^2 (\alpha  \xi  A' (\xi ^2 B^2 (32 \pi  P_{rc} ((6 N^3-a \xi ^3) \psi_{o}^n (n (\alpha -\alpha_{1}) \psi_{o}\notag\\
&&+\alpha  (-n)+\alpha_{1} n+1)-3 v)+a \alpha  \xi  (3 N^2+32 \pi  \xi ^2))-12 \alpha  N^2 (a \xi ^3\notag\\
&&+2 N^3))+A (12 \alpha ^2 N^2 (4 N^3-a \xi ^3)+\xi ^2 B (2 \alpha  \xi  B' (32 \pi  P_{rc} ((6 N^3\notag\\
&&-a \xi ^3) \psi_{o}^n (n (\alpha -\alpha_{1}) \psi_{o}+\alpha  (-n)+\alpha_{1} n+1)-3 v)+a \alpha  \xi  (3 N^2\notag\\
&&+32 \pi  \xi ^2))+B (3 a \xi  (32 \pi  \alpha ^2 \xi ^2+\alpha ^2 N^2-\frac{4 \pi  P_{rc}}{n+1})+4 \alpha  (3 \alpha  N^4 q'\notag\\
&&-8 \pi  n \xi  P_{rc} \psi_{o}^{n-1} ((a \xi ^3-6 N^3) \psi_{o}' ((n+1) (\alpha -\alpha_{1}) \psi_{o}+\alpha  (-n)\notag\\
&&+\alpha_{1} n+1)-3 a \alpha  \alpha_{1} n \xi ^2 (\psi_{o}-1) \psi_{o}))))))\big]+12 N \xi  q (-4 N^6 q'\notag\\
&&+N^3 \xi ^3 B^2 A'+A B (2 N^3 \xi ^3 B'+B (4 q' (-3 N^6-\frac{2 \pi  \xi ^3 P_{rc}}{\alpha ^2 (n+1)})\notag\\
&&-N^3 \xi ^2)))\Big]=0.
\end{eqnarray}
Now Eqs. $(\ref{43}, \ref{44}, \ref{55})$ form a system of ordinary DEs having three variables  v, $\psi_{o}$ and $\Delta$. This system of ordinary DEs is solved numerically and its solution is described graphically. Figures ($1$-$3$) depicted the behavior of $v$, $\psi_{o}$ and $\Delta$ for $\alpha_{1}=.5$, $\alpha=.5$, $n=.5$ and $a=5$.
\begin{figure}[!htbp]\label{fig1} 
\centering
\includegraphics[width=100mm]
{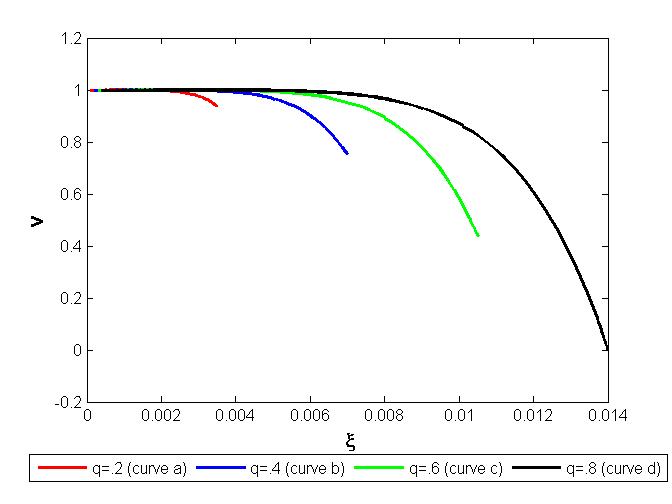}
\hspace{200mm}
\caption{Graphs between $\xi$ and $v$}
\end{figure}

\begin{figure}[!htbp]\label{fig2} 

{\includegraphics[width=90mm]
{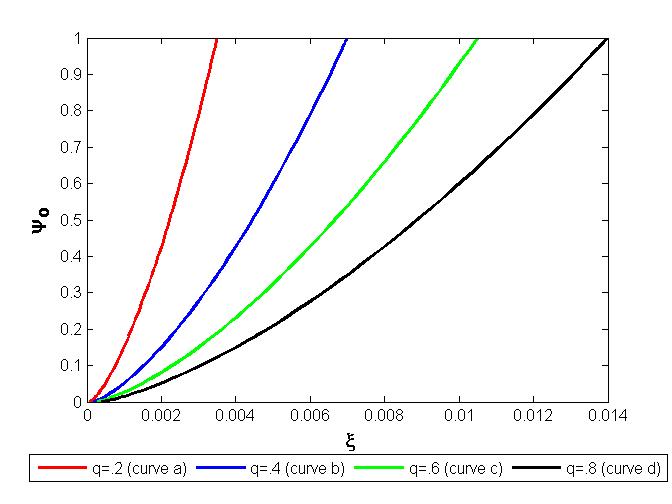}}
\centering
\caption{Graphs between $\xi$ and $\psi_{o}$}
\end{figure}
\begin{figure}[!htbp]\label{fig3} 
\centering
{\includegraphics[width=90mm]{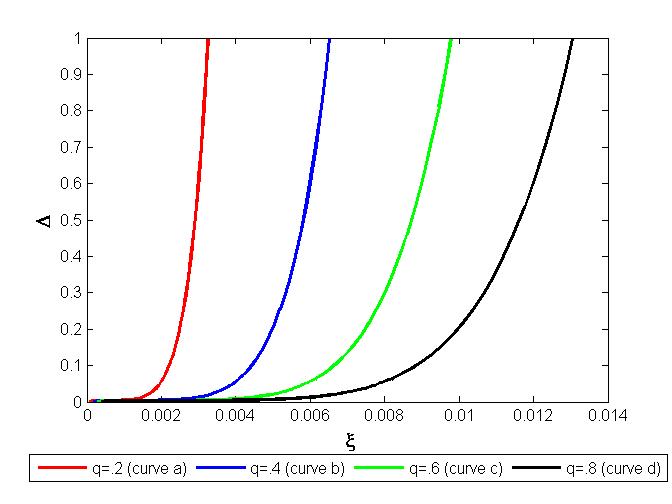}}
\hspace{200mm}
\caption{Graphs between $\xi$ and $\Delta$}
\end{figure}
\FloatBarrier
\subsection{Case No.2}
 CF for case $2$ will be read the  as
\begin{eqnarray}\label{56}\notag
&&4 \pi  (\Delta'+2 B^2 (\Delta A'+A \Delta')+4 A B \Delta B')-\frac{1}{24 N^3 \xi ^5}\Big[24 N^2 q^2\notag\\
&& (2 A B (B (\frac{\pi  (a+1) \xi ^3 P_{rc}}{\alpha ^2 (n+1)}+6 N^5)-3 N^5 \xi  B')+N^5 (4-3 \xi  B^2 A'))\notag\\
&&+\frac{1}{\alpha ^2}\big[\xi ^2 (\alpha  \xi  A' (\xi ^2 B^2 (a \alpha  \xi  (3 N^2+32 \pi  \xi ^2)-32 \pi  P_{rc} ((a \xi ^3-6 N^3) \psi ^n+3 v))\notag\\
&&-12 \alpha  N^2 (a \xi ^3+2 N^3))+A (12 \alpha ^2 N^2 (4 N^3-a \xi ^3)+\xi ^2 B (2 \alpha  \xi  B'\notag\\
&& (a \alpha  \xi  (3 N^2+32 \pi  \xi ^2)-32 \pi  P_{rc} ((a \xi ^3-6 N^3) \psi ^n+3 v))+B (3 a \xi  (32 \pi  \alpha ^2 \xi ^2\notag\\
&&+\alpha ^2 N^2-\frac{4 \pi  P_{rc}}{n+1})+4 \alpha  (3 \alpha  N^4 q'-8 \pi  n \xi  P_{rc} (a \xi ^3-6 N^3) \psi ^{n-1} \psi ')))))\big]\notag\\
&&+12 N \xi  q (-4 N^6 q'+N^3 \xi ^3 B^2 A'+A B (2 N^3 \xi ^3 B'+B (4 q' (-3 N^6\notag\\
&&-\frac{2 \pi  \xi ^3 P_{rc}}{\alpha ^2 (n+1)})-N^3 \xi ^2)))\Big]=0.
\end{eqnarray}
Eqs. $(\ref{49}, \ref{50}, \ref{56})$ constitute a system of ordinary DEs containing three variables $\Delta$, $v$ and $\psi$. This system of ordinary DEs is numerically solved  and its solution is explained through graphs in Figures ($4$-$6$) for $\alpha_{1}=.5$, $\alpha=.5$, $n=.5$ and $a=5$.

\begin{figure}[!htbp]\label{fig4} 
\centering
{\includegraphics[width=80mm]{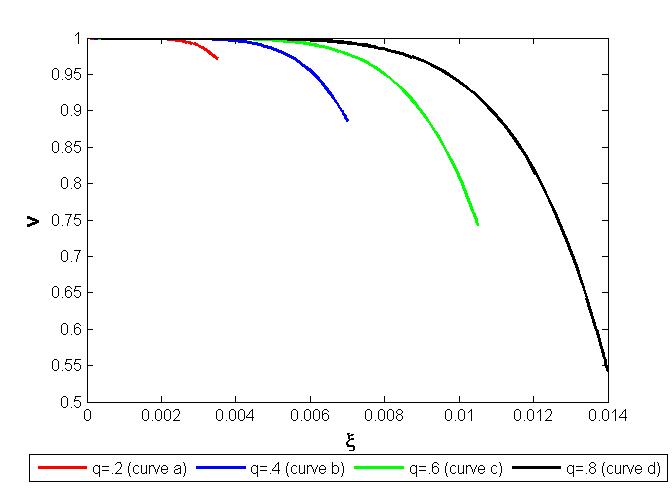}}
\hspace{200mm}
\caption{Graphs between $\xi$ and $v$}
\end{figure}
\begin{figure}[!htbp]\label{fig5} 
\centering
{\includegraphics[width=90mm]{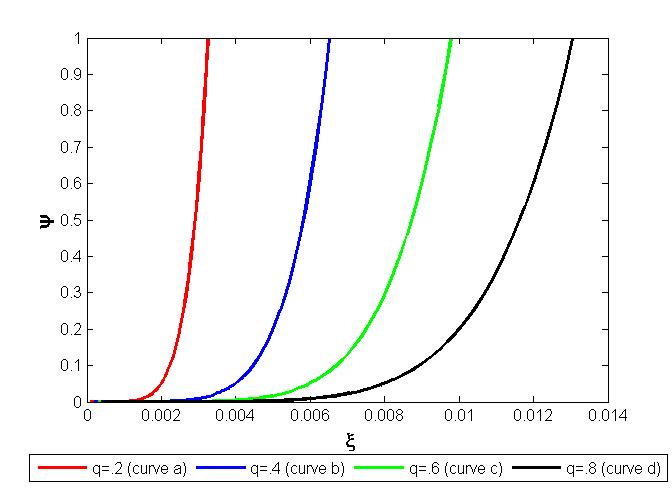}}
\hspace{200mm}
\caption{Graphs between $\xi$ and $\psi$}
\end{figure}
\begin{figure}[!htbp]\label{fig6} 
\centering
{\includegraphics[width=90mm]{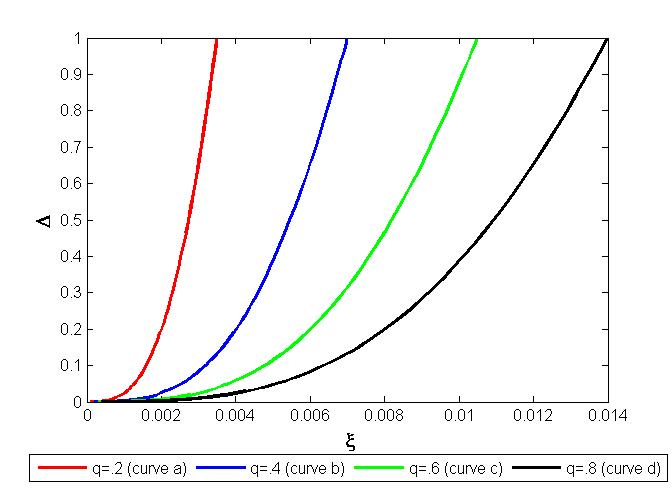}}
\hspace{200mm}
\caption{Graphs between $\xi$ and $\Delta$}
\end{figure}
\FloatBarrier
\section{Conclusion}
In the recent past the idea of GPEoS $[27-33, 37, 38,39]$ and complexity factor $[33-39]$ has been widely used to study the different physical features of self gravitating stellar object. In this paper we applied these two ideas to discuss some physical properties $(v, \psi_{o},\psi,\Delta)$ of of these objects. For this purpose an anisotopic fluid distribution with static  cylindrical symmetry is used to establish a general framework for the development of modify form of LEs in the presence of electromagnetic field. Basic field equations are used to establish TOV equation.  The C-energy with charge is put on to develop a expression for mass function. Curvature and  Weyl tensors are brought into play to calculate the structure scalars. These structure scalars are used to define vanishing CF. We applied assumptions Eqs. $(\ref{41}, \ref{42})$ to establish the LEes for static charged cylindrical fluid distribution for two cases $\textbf{(1)}$ mass density and $\textbf{(2)}$ energy density to study the physical characteristics of GPs. The energy conditions under the influence of electromagnetic effect have also been studied for these two cases under same the assumptions. These two set of LEes $(\ref{43}, \ref{44})$ and $(\ref{49}, \ref{50})$ led us to  the two system of three ordinary DEs with CF. These system of DEs numerically solved  and discussed below. \\

The solutions of set of DEs. $(\ref{43}, \ref{44}, \ref{55})$ and $(\ref{43}, \ref{44},\ref{56})$ are set out in the Figs. $(\textbf{1-3})$ and Figs. $(\textbf{4-6})$ for case $(\textbf{1})$ and case $(\textbf{2})$ respectively. For different values of parameters these solutions exhibit the behavior of variables $v$, $\psi_{o}$, $\Delta$ for case $(\textbf{1})$ and $v$, $\psi$, $\Delta$ for case $(\textbf{2})$.\\
The curves of Figs. $(\textbf{1})$ and $(\textbf{4})$ in both cases show that the value of $v$ have maximum value at center of the stellar structure for different value of charge and then value of $v$ decreases with the increase of radius. It is also observed that value of $v$ gradually becomes zero with increase of charge.\\
In Figs. $(\textbf{2})$ and $(\textbf{6})$ curves indicate pattern of $\psi_{o}$ in case $(\textbf{1})$ and $\Delta$ in case $(\textbf{2})$ respectively. Curves of both Figs. are similar which indicate minimum value of $\psi_{o}$ and $\Delta$ at center of stellar structure and uniform increase towards boundary surface. It can also be seen that these variables attain maximum value with gradually rise in charge in increasing direction of radius.\\
Fig. $(\textbf{3})$ and Fig. $(\textbf{5})$ display the pattern of $\Delta$ for case $(\textbf{1})$ and $\psi$  for case $(\textbf{2})$. It can be observed that behavior of Figs. $(\textbf{2})$, $(\textbf{6})$ and Figs. $(\textbf{3})$, $(\textbf{5})$ are all most same except the curves of Fig. $(\textbf{3})$, and Fig. $(\textbf{5})$ which show sudden increase with increase of radius for smaller value of charge. Hence the main purpose of this work is to represent a general framework for the development of modify form of LEes for charged generalized polytropes in the context of cylindrical symmetry.

}

\begin{thebibliography} {50}
\bibitem{28}  W. B. Bonnor, The mass of a static charged sphere,The mass of a static charged sphere, Zeit. Phys. \textbf{160} (1960) 59

\bibitem{29}  H. Bondi, The Contraction of Gravitating Spheres, Proc. R. Soc. Lond. A \textbf{281} (1964) 39

\bibitem{50} K. S. Thorne, Energy of infinitely long, cylindrically symmetric systems in general relativity5, Phys. Rev. \textbf{138}, (1965), B251

\bibitem{27} J. D. Bekenstein, Hydrostatic equilibrium and gravitational collapse of relativistic charged fluid balls, Phys. Rev. D \textbf{4}, (1971) 2185

\bibitem{51} M. K. Mak and T. Harko, Quark stars admitting a one-parameter group of conformal motions, Inter. J. M. Phys \textbf{13} (2004) 149

\bibitem{31} P. M. Takisa and S. D. Maharaj, Some charged polytropic models,  Astrophys. Space Sci. \textbf{45} (2013) 1951

\bibitem{34} M. Sharif and S. Sadiq, Electromagnetic effects on cracking of anisotropic polytropes Eur. Phys. J. C \textbf{76} (2016) 568
\bibitem{39} J. H. Lane, On the Theoretical Temperature of the Sun; under the Hypothesis of a Gaseous Mass maintaining its Volume by its Internal Heat, and depending on the Laws of Gases as known to Terrestrial Experiment, Am. J. Sci. Arts. \textbf{50} (1870) 148
 
 \bibitem{1} S. Chandrasekhar, An introduction to the Study of Stellar Structure, University of Chicago, Chicago, (1939)
 
\bibitem{2}  R. F. Tooper, General relativistic polytropic fluid spheres, Astrophys. J. \textbf{140} (1964) 434

\bibitem{3} R. F. Tooper, Adiabatic fluid spheres in general relativity, Astrophys. J. \textbf{142} (1965) 1541

\bibitem{4} S. A. Kaplan and G. A. Lupanov, The relativistic instability of polytropic spheres, Soviet Astronomy  \textbf{9}  (1965) 233

\bibitem{40}  J. J. Managhan and W.  Roxburgh, The Structure of rapidly rotating polytropes, Mon. Not. R. astr.Soc. \textbf {131} (1965) 13
 
\bibitem{5} W. J. Kaufmann, Polytropic spheres in general relativity astrophys. J. \textbf{72},  (1967) 754

\bibitem{6}  F. Occhionero, Rotationally distorted polytropes second order approximation $n\geq2$,  Mem. Soc. Astron. Ital. \textbf{38} (1967) 3310

\bibitem{7}A. Kovetz, Slowly rotating polytropes,  Astrophys. J. \textbf{154}, 999 (1968)
\bibitem{8} G. P. Horedt, Weakly distorted polytropes under external pressure,  Astron. Astrophys. \textbf {23}  (1973) 303

\bibitem{9} J. P. Sharma, Relativistic spherical polytropes:
an analytical approach, Gen. Relat. Gravit. \textbf{13} (1981) 663

\bibitem{41}  M. Singh and G. Singh, The structure of the tidally and rotationally distorted polytropes, Astrophys. Space Sci. \textbf{96} (1983) 313


\bibitem{11}  G. P. Horedt, Physical characteristics of n-dimensional radially- symmetric polytropes, Astrophys. Space Sci.\textbf{133} (1987) 81

\bibitem{12}  S. C. Pandey, et al.,Relativistic polytropic spheres in general relativity, Astrophys. Space Sci. \textbf{180} (1991) 75

\bibitem{42}  A. W. Hendry, A polytropic model of the sun, Am. J. Phys. \textbf{61} (1993) 906

\bibitem{14}  L. Herrera and W. Barreto, Evolution of relativistic polytropes in the post–quasi–static Regime, Gen. Rel. Grav. \textbf{36} (2004) 127 
 
\bibitem{16} L. Herrera and W. Barreto, General relativistic polytropes for anisotropic matter: The general formalism
and applications Phys. Rev. D \textbf{88} (2013) 084022

\bibitem{17}  L. Herrera, et al., Conformally flat polytropes for anisotropic matter Gen. Relat. Gravit. \textbf{46} (2014)  1827
\bibitem{18} L. Herrera, et al., Cracking of general relativistic anisotropic polytropes, Phys. Rev. D \textbf{93} (2016)  024047

\bibitem{19} M. Azam, et al., Study of polytropes with generalized polytropic equation of state, Eur. Phys. J. C \textbf{76} (2016) 315


\bibitem{35} M. Azam and S. A. Mardan, Cracking of anisotropic cylindrical polytropes, Eur. Phys. J. C \textbf{77}  (2017) 113

\bibitem{43}  S. A. Mardan, et al., New classes of anisotropic models with generalized polytropic equation of state, Eur. Phys. J. C \textbf{78} (2018)  516

\bibitem{44}  S. A. Mardan, et al., New classes of generalized anisotropic polytropes pertaining radiation density, Eur. Phys. J. Plus \textbf{134}, (2019)  242

\bibitem{45}  S. A. Mardan, et al., New models of charged anisotropic polytropes with radiation density Eur. Phys. J. Plus \textbf {135} (2020) 3

\bibitem{46}  S. A. Mardan, et al., Impact of generalized polytropic equation of state on charged anisotropic polytropes Eur. Phys. J. C \textbf{80}, (2020) 119
   
\bibitem{21}  L. Herrera, New definition of complexity for self–gravitating fluid distributions: The spherically
symmetric, static case, Phys. Rev. D. \textbf{97} (2018) 044010

\bibitem{47} G. Abbas and H. Nazar, Complexity factor for static anisotropic self-gravitating source in f (R) gravity Eur. Phys. J. C \textbf{78} (2018) 510


\bibitem{36}  M. Sharif and I. I. Butt, Complexity factor for static cylindrical system, Eur. Phys. J. C \textbf{78} (2018) 850

\bibitem{48}  M. Sharif and I. I. Butt, Electromagnetic effects on complexity factor for static cylindrical system Chin. J. Phys.  \textbf{61} (2019) 238

\bibitem{32} S. Khan, et al., Framework for generalized polytropes with complexity factor Eur. Phys. J. \textbf{1037},  (2019) 79

\bibitem{49}  S. Khan, et al., Study of charged generalized polytropes with complexity factor Eur. Phys. J. Plus \textbf{136} (2021) 404

\bibitem{52} S. Khan, et al., Study of generalized cylindrical polytropes with complexity factor Eur. Phys. J. C \textbf{81} (2021) 831


\bibitem{37} M. Sharif and G. Abbas, AstoPhys. Space Sci \textbf{335}, 515 (2011)



\bibitem{24}  L. Bel, Inductions électromagnétique et gravitationnelle, \textit{Ann. Inst. H Poincar$\acute{e}$}  \textbf{17} (1961) 37

\bibitem{26}  L. Herrera, et al., 
Structure and evolution of self-gravitating objects and the orthogonal splitting of the Riemann tensor Phys. Rev. D \textbf{79} (2009) 064025

\bibitem{33}  L. Herrera, et al., Role of electric charge and cosmological constant in structure scalars, Phys. Rev. D \textbf{84}, (2011) 107501

\bibitem{15}  L. Herrera and W. Barreto, Newtonian polytropes for anisotropic matter: General framework and applications Phys. Rev. D \textbf{87} (2013) 087303
\end{thebibliography}
\end{document}